\documentclass{article}

\usepackage{arxiv}

\usepackage[utf8]{inputenc} 
\usepackage[T1]{fontenc}    
\usepackage{hyperref}       
\usepackage{url}            
\usepackage{booktabs}       
\usepackage{amsfonts}       
\usepackage{nicefrac}       
\usepackage{microtype}      
\usepackage{lipsum}		    
\usepackage{graphicx}
\usepackage{natbib}
\usepackage{doi}

\title{Nie pozwól algorytmom rządzić Twoim koszykiem: systemy rekomendacyjne w dobie Omnibusa
}
\date{}

\author{ Mikołaj Morzy \\
	Politechnika Poznańska \\
	ul. Piotrowo 2, 60-965 Poznań, Poland \\
	\texttt{mikolaj.morzy@put.poznan.pl} \\
	\And
	Mirosław Sobieraj \\
	Trol Intermedia sp. z o.o.\\
	ul. Poznańska 56, 64-510 Wronki, Poland \\
	\texttt{m.sobieraj@trol.pl} \\
	\AND
	Sebastian Sikora \\
	Trol Intermedia sp. z o.o.\\
	ul. Poznańska 56, 64-510 Wronki, Poland \\
	\texttt{s.sikora@trol.pl} \\
}

\renewcommand{\headeright}{Raport techniczny}
\renewcommand{\undertitle}{Raport techniczny}
\renewcommand{\shorttitle}{Systemy rekomendacyjne w dobie Omnibusa}

\begin{document}
\maketitle

\begin{abstract}
Dyrektywa Omnibus stanowi istotną część Nowego Ładu dla Konsumentów (ang. \emph{New Deal for Consumers}) Unii Europejskiej. Dyrektywa wprowadza nowe regulacje w handlu, w tym e-commerce, których głównym celem jest zwiększenie przejrzystości, uczciwości i ochrony konsumentów. Autorzy krytycznie zwracają uwagę na istotne zaniedbanie w dyrektywie Omnibus, jakim jest brak uwzględnienia systemów rekomendacyjnych. Silniki rekomendacyjne mogą być źródłem potencjalnie szkodliwych praktyk uderzających w konsumentów, stąd niezbędne jest rozszerzenie dyrektywy. Propozycje przedstawione w niniejszym artykule obejmują wprowadzenie etycznego nadzoru nad systemami rekomendującymi, aby zminimalizować ryzyko negatywnych skutków ich rekomendacji, a także jasne wyjaśnienie kryteriów, na podstawie których dokonywane są rekomendacje - analogicznie do rankingów wyników wyszukiwania.

\end{abstract}


\section{Wprowadzenie}
Oczekuje się, że w nadchodzących latach zakupy spożywcze online będą nadal rosły i ulegną znaczącym zmianom. Według przewidywań, sprzedaż online produktów spożywczych do 2030 roku ma osiągnąć 26\% w Wielkiej Brytanii, 25\% w Holandii, 23\% we Francji, 11\% w Niemczech i 10\% w Turcji. Zachowania i preferencje klientów będą się zmieniać, ponieważ cyfrowi tubylcy i wszystkie grupy wiekowe coraz częściej korzystają z zakupów online. Godnym uwagi trendem jest rozwój wielokanałowego handlu detalicznego, w którym konsumenci płynnie przełączają się między sklepami fizycznymi a platformami internetowymi. W Polsce do końca 2022 roku sprzedaż produktów spożywczych osiągnęła 1,5\%, a do 2027 roku ten wskaźnik wzrośnie do 2,7\%. Wzrost będzie jednak skoncentrowany głównie w obszarach miejskich. W największych aglomeracjach wskaźnik ten może przekroczyć 10\% pod koniec bieżącej dekady, podczas gdy na terenach wiejskich i w mniejszych miastach możliwości zakupu świeżych produktów spożywczych online mogą być ograniczone. Z drugiej strony, widać, że z każdym rokiem coraz więcej konsumentów robi większość swoich zakupów spożywczych online. Dlatego można spodziewać się, że pomimo niskiego obecnego wskaźnika sprzedaży online w ogólnym obrocie, możemy w niedalekiej przyszłości oczekiwać rejonów, w których sprzedaż online stanowić będzie więcej niż 20\%. 

Polacy wykazują wyższą częstotliwość robienia zakupów spożywczych niż ich zachodni odpowiednicy, dokonując ich średnio 2-3 razy w tygodniu, w porównaniu do 1-2 razy na Zachodzie. Około 40\% Polaków przyznaje się do codziennych zakupów spożywczych. Ta różnica między Polską i Zachodem jest częściowo wynikiem preferencji polskich konsumentów dla świeżych produktów. Dodatkowo, dostępność sklepów typu convenience i dyskontów sprzyja większej częstotliwości zakupów. W przewidywaniach zakłada się, że to dyskonty i sklepy typu convenience będą najdynamiczniej się rozwijać, na niekorzyść supermarketów, hipermarketów oraz sklepów tradycyjnych. Trend convenience znajduje w Polsce silne podłoże. Głównymi czynnikami skłaniającymi klientów do korzystania na przykład z samoobsługowych kas, są wygoda (76\%) oraz oszczędność czasu (73\%). Te same powody przyciągają klientów do zakupów online, co wpływa na rosnącą popularność sklepów typu \emph{convenience}, które oferują szybką obsługę, dostępność i asortyment dobrany do potrzeb klienta. Jeszcze jednym krokiem do przodu dla sklepów \emph{convenience} jest rozwój sklepów autonomicznych, których Polska posiada najwięcej w Europie, a aż 70\% Polaków jest zainteresowanych skorzystaniem z takiej opcji.

Unijna dyrektywa Omnibus została przedstawiona jako kompleksowa odpowiedź na rosnące obawy dotyczące ochrony konsumentów i zapewnienia przejrzystości na rynku tradycyjnym i cyfrowym. Dyrektywa, której celem jest wzmocnienie praw konsumentów, obejmuje kilka kluczowych obszarów handlu detalicznego online - w tym środki mające na celu zwalczanie ukrytych reklam, kontrolę fałszywych recenzji oraz zwiększoną przejrzystość spersonalizowanych cen. Ma ona na celu zapewnienie solidnych zabezpieczeń przed nieuczciwymi praktykami handlowymi, wspierając zaufanie wśród europejskich konsumentów. Jednak w ewoluującym krajobrazie handlu detalicznego online istnieje zauważalna luka w tej zbroi ochronnej - regulacja internetowych silników rekomendacyjnych. Pomimo ich powszechnego wpływu i wykorzystania, silniki te --- zasilane złożonymi algorytmami w celu dostosowania indywidualnych doświadczeń konsumentów --- nie są wyraźnie uwzględnione w dyrektywie, pozostawiając ważny obszar ochrony konsumentów i przejrzystości bez dedykowanych ram prawnych.

\section{Czym jest dyrektywa Omnibus?}

Dyrektywa Omnibus, przyjęta przez Parlament Europejski i Radę Europy w listopadzie 2019 r., to kompleksowy pakiet środków mających na celu wzmocnienie ochrony konsumentów w całej Unii Europejskiej. Dyrektywa opiera się na istniejącym prawodawstwie w zakresie ochrony konsumentów i wprowadza szereg nowych przepisów w celu sprostania wyzwaniom związanym z erą cyfrową i zmieniającym się krajobrazem konsumenckim.

Kluczowe obszary zainteresowania obejmują zwiększenie przejrzystości i wymogów informacyjnych dla przedsiębiorców, wzmocnienie praw konsumentów w odniesieniu do umów cyfrowych i internetowych platform handlowych oraz poprawę mechanizmów egzekwowania prawa w celu zwalczania nieuczciwych praktyk handlowych i naruszeń praw konsumentów. Dyrektywa wprowadza również środki mające na celu ochronę konsumentów przed wprowadzającymi w błąd i agresywnymi praktykami marketingowymi, nieuczciwymi warunkami umów oraz niebezpiecznymi produktami i usługami.

Dyrektywa OMNIBUS opiera się na trzech filarach: przejrzystości, uczciwości i kontroli konsumenckiej. Ideą wymagania przejrzystości jest zapewnienie konsumentom dostępu do rzetelnych i pełnych informacji o produktach i usługach oferowanych przez przedsiębiorców. Dyrektywa nakłada na przedsiębiorców szereg obowiązków informacyjnych, które mają pomóc konsumentom w podejmowaniu świadomych decyzji zakupowych. W sklepach internetowych przedsiębiorcy muszą ujawniać następujące informacje:

\begin{itemize}
    \item cenę produktu lub usługi, w tym wszelkie opłaty dodatkowe, takie jak koszty dostawy lub montażu,
    \item cenę poprzednią, jeśli produkt lub usługa jest objęta promocją,
    \item główne cechy produktu lub usługi, w tym jego skład, właściwości, sposób użytkowania i konserwacji,
    \item terminy i warunki gwarancji oraz rękojmi,
    \item informacje o sposobie składania reklamacji,
    \item informacje o prawach konsumenta, takich jak prawo do odstąpienia od umowy w terminie 14 dni.
\end{itemize}

Przedsiębiorcy są zobowiązani do jasnego i jednoznacznego informowania konsumentów o warunkach promocji, takich jak okres obowiązywania promocji, wysokość rabatu lub gratisu, a także o ograniczeniach, które mogą dotyczyć promocji. Dodatkowo, przedsiębiorcy są zobowiązani do ujawniania tych ofert, które są wynikiem partnerstw handlowych lub są tzw. miejscami sponsorowanymi, których widoczność nie musi być jednoznacznie wynikiem wyszukiwania wykonanego przez klienta sklepu internetowego. 

Centralnym celem dyrektywy Omnibus jest stworzenie bardziej przejrzystego środowiska dla zakupów online. W tym celu wprowadzone zostały regulacje, które zobowiązują przedsiębiorców do informowania konsumentów o tzw. rankingowaniu produktów, który we wprowadzonych przepisach w Polsce jest określany mianem plasowania. Przepisy krajowe określają ranking jako proces nadawania produktom konkretnego stopnia widoczności lub przypisywanie wagi wynikom wyszukiwania przez usługi oferujące funkcje wyszukiwania. Zgodnie z definicją podaną w dyrektywie Omnibus, ranking może wynikać m.in. z zastosowania algorytmicznego sekwencjonowania, ocen lub opinii, wyróżniających cech bądź ich kombinacji. Polski legislator nie skopiował listy zaproponowanej w dyrektywie Omnibus, stwierdzając, że plasowanie jest mechanizmem niezależnym od rodzaju użytej technologii.

Ciekawym wymogiem wprowadzonym przez dyrektywę Omnibus są nowe zasady które mają zastosowanie do recenzji produktowych. Zostały one określone w zaktualizowanym Załączniku I do dyrektywy. Zgodnie z tymi zasadami, za wprowadzającą w błąd praktykę handlową uznaje się zlecanie przez przedsiębiorcę (np. internetową platformę handlową lub sprzedawcę na takiej platformie) recenzji własnych towarów lub usług, lub towarów lub usług konkurenta, bez wyraźnego wskazania tego faktu. Ponadto aktualizacja wyjaśnia, że za nieuczciwą praktykę uznaje się manipulowanie przez przedsiębiorcę opiniami użytkowników na temat jego towarów lub usług, na przykład poprzez publikowanie wyłącznie pozytywnych recenzji lub zmianę ich treści. Ogólnym celem tych zasad jest zapewnienie przejrzystości konsumentom, którzy polegają na recenzjach przy podejmowaniu decyzji zakupowych, zapewniając im dokładne, uczciwe przedstawienie produktu, usługi lub dostawcy. Wymagając od recenzentów ujawnienia wszelkich powiązań, wynagrodzeń lub istniejących relacji z recenzowanym podmiotem, konsumenci mogą krytycznie ocenić wiarygodność i rzetelność przedstawionej im recenzji. 

Dyrektywa Omnibus wspiera uczciwość w handlu internetowym i stacjonarnym poprzez wprowadzenie szeregu zakazów dotyczących wprowadzających w błąd praktyk. Wprowadzające w błąd praktyki to takie, które okłamują konsumentów co do istotnych cech produktu lub usługi, takich jak cena, jakość, właściwości lub warunki gwarancji. Dyrektywa Omnibus zakazuje między innymi: 

\begin{itemize}
    \item używania nieprawdziwych lub wprowadzających w błąd informacji o produkcie lub usłudze,
    \item ukrywania lub pomijania istotnych informacji o produkcie lub usłudze,
    \item używania nieuczciwej porównawczej reklamy,
    \item używania agresywnych praktyk sprzedażowych.
\end{itemize}

Dyrektywa wprowadza także zakaz użycia technik, które wykorzystują ograniczenia poznawcze konsumentów w celu skłonienia ich do podjęcia decyzji, których w przeciwnym razie by nie podjęli. Przykładami takich technik może być utrudnianie konsumentom dokonania świadomej decyzji zakupowej, korzystanie z mechanizmów uzależniających których celem jest skłonienie do niepotrzebnych zakupów, lub wykorzystywanie danych osobowych w celu niechcianej personalizacji. W kontekście sklepu internetowego wyrazem zakazanych praktyk może być ogłoszenie promocji na produkt, który jest zawsze w “promocyjnej” cenie, użycie czcionki o zbyt małej wielkości w warunkach umowy, czy wyświetlenie okienka typu pop-up z reklamą w momencie, gdy konsument chce zamknąć stronę internetową.

Głównym celem dyrektywy Omnibus jest zapewnienie konsumentom równych szans w podejmowaniu świadomych decyzji zakupowych. Oczywiście cel ten jest w dużej mierze sprzeczny z celami sprzedawców, których interesuje zwiększenie wolumenu sprzedaży. Szczególnie jest to widoczne na przykładzie sklepów internetowych wyposażonych w moduły sztucznej inteligencji. Jednym z obszarów, w których wpływ dyrektywy będzie w najbliższym czasie najbardziej widoczny (poza jawnym raportowaniem najniższej ceny produktu w przeciągu ostatnich 6 miesięcy), będą wyniki wyszukiwania. Dyrektywa Omnibus nakazuje przejrzystość w wyświetlaniu wyników wyszukiwania na platformach internetowych, takich jak witryny handlu elektronicznego. Zgodnie z dyrektywą, sklepy internetowe są zobowiązane do przedstawienia głównych parametrów określających ranking wyników wyszukiwania. W praktyce oznacza to wyjaśnienie kupującym, w jaki sposób zapytania i rankingi są przetwarzane i na jakich czynnikach się opierają --- czy jest to popularność, poprzednie zakupy czy opinie konsumentów. Celem tej zasady w ramach dyrektywy Omnibus jest zapewnienie, że konsumenci nie są manipulowani przez ukryte algorytmy i mogą dokonywać świadomych wyborów dotyczących zakupów w oparciu o otwarcie dostępne i zrozumiałe kryteria. W praktyce oznacza to, że sklepy internetowe nie będą mogły prezentować wyników wyszukiwania według “trafności” bez precyzyjnego zdefiniowania, czym owa trafność jest. Dziś pod pojęciem “trafności” kryje się najczęściej skomplikowana formuła biorąca pod uwagę dopasowanie produktu do zapytania klienta, “opłacalność” produktu rozumiana jako stosunek liczby wyświetleń do liczby zamówień, stan magazynowy, marża, prawdopodobieństwo zakupu produktu przez klienta, wzorzec wynikający z wcześniejszej historii interakcji klienta ze sklepem, i zapewne jeszcze tuzin innych czynników. Bez wątpienia tak zdefiniowana “trafność” nie spełnia postulatu przejrzystości.

\section{Studium przypadku: Inteligentny System Rekomendacyjny}

W latach 2018-2020 Autorzy niniejszego artykułu pracowali nad zbudowaniem Inteligentnego Systemu Rekomendacyjnego (ISR) dla rynku e-Commerce. Celem projektu było stworzenie zaawansowanej technologii wykorzystującej sztuczną inteligencję i analizę danych w celu dostarczenia spersonalizowanych rekomendacji produktowych dla użytkowników e-sklepów. System ten miał na celu zwiększenie efektywności sprzedaży i poprawę doświadczeń zakupowych. Przez okres trwania projektu, zespół badawczy skupił się na opracowaniu algorytmów uczenia maszynowego, które umożliwiłyby dokonanie analizy dużych zbiorów danych, uwzględniając historię zakupów, preferencje oraz zachowania użytkowników. Dzięki temu, system rekomendacji posiada zdolność do precyzyjnego dopasowywania produktów do indywidualnych potrzeb klientów. Realizacja projektu obejmowała kilka etapów, począwszy od badań i rozwoju, poprzez testowanie prototypów, aż po wdrożenie systemu w środowisku rzeczywistym. Współpraca z przedsiębiorstwami i instytucjami naukowymi odegrała kluczową rolę w udoskonaleniu technologii i zapewnieniu jej praktycznej użyteczności. Projekt, który zakończył się sukcesem, nie tylko przyczynił się do rozwoju nowoczesnych technologii w dziedzinie e-commerce, ale także stanowił istotny wkład w rozwój badań naukowych w obszarze sztucznej inteligencji i analizy danych. Wyniki projektu otworzyły nowe możliwości dla przyszłych innowacji i zastosowań w różnych sektorach gospodarki.

Zaprojektowany ISR bazuje na koncepcji atomowego zdarzenia w trakcie interakcji klienta ze e-sklepem. Zdarzeniem może być akcja wykonana przez klienta (włożenie produktu do koszyka, wejście na stronę kategorii, zakup produktu), może to być cecha wynikająca z zachowania klienta (pozostanie przez ponad 30 sekund na stronie produktu), może to być zdarzenie z przeszłości (w ciągu ostatnich 6 miesięcy klient zakupił produkt w tej samej kategorii), wreszcie może to być statyczna cecha klienta (klient zaznaczył, że jest wegetarianinem) albo produktu (produkt zawiera olej palmowy). Silnik rekomendacji przypisuje poszczególnym zdarzeniom różną wagę w zależności od globalnej konfiguracji, a następnie wykorzystuje opracowany przez zespół zaawansowany algorytm zaimplementowany w środowisku silnika rekomendacyjnego w celu wytworzenia rekomendacji. Opracowany algorytm to nowy typ algorytmu rekomendacyjnego opartego na filtrowaniu kolaboracyjnym, bazujący na algorytmie Correlated Cross-Occurrence (CCO), który może wykorzystywać dane z wielu różnych źródeł w celu lepszego dostrajania i tworzenia rekomendacji. W przeciwieństwie do metod bazujących na faktoryzacji macierzy, algorytm CCO jest w stanie przyjąć dowolną liczbę działań użytkownika, zdarzeń, danych profilowych i informacji kontekstowych. Obsługuje również reguły biznesowe, które filtrują listy rekomendacji i dopasowują owe listy do specyficznych polityk. Polityki takie regulują rekomendacje pod kątem grupy użytkowników (rekomendacje dla stałych klientów), specyficznych warunków czasowych (zmiana rekomendacji na tydzień przed Walentynkami), czy uregulowań prawnych (zakaz rekomendowania alkoholu osobom nieletnim). Opracowany algorytm jest hybrydowym silnikiem filtrowania kolaboracyjnego i rekomendacji opartych na treści.

To właśnie podczas projektowania reguł biznesowych kontrolujących zachowania silnika rekomendacyjnego odkryliśmy, że dyrektywa Omnibus, którą równolegle zespół projektowy wdrażał w oprogramowaniu sklepu internetowego, całkowicie pomija kwestię rekomendacji. Do silnika rekomendacyjnego nie stosowały się ani zalecenia dotyczące wyników wyszukiwania, ani przepisy regulujące kwestie reklam produktowych, ani zasady zdefiniowane dla cen czy recenzji produktowych. Można powiedzieć, że dyrektywa Omnibus całkowicie pominęła cały obszar rekomendacji, mimo, że silnik rekomendacyjny też może być mechanizmem wdrażania złych i szkodliwych z punktu widzenia klienta praktyk. Rozważmy kilka przykładowych scenariuszy.

\begin{itemize}
    \item Klient sklepu internetowego w informacjach profilowych zaznacza, że jest wegetarianinem i chce unikać produktów wysokocukrowych. Mimo to, silnik rekomendacji może umieszczać w ramkach rekomendacyjnych obrazy produktów odzwierzęcych (rekomendacja narusza światopogląd klienta) lub produktów o wysokiej zawartości cukru (rekomendacja szkodliwie wpływająca na zdrowie klienta).
    \item Klient sklepu zaznacza, że jest osobą pełnoletnią. System rekomendacyjny zaczyna rekomendować mocne alkohole jako produkty o dużym marginesie zysku, nawet jeśli może to wspomagać szkodliwe nałogi klienta lub nie mieć żadnego związku z bieżącym kontekstem zakupowym klienta.
    \item Reguła biznesowa priorytetowo traktuje produkty określonego sprzedawcy, mimo, że sklep posiada tańsze zamienniki towarów, rekomendacja podyktowana jest jedynie interesem sprzedawcy, który może liczyć na korzyść finansową związaną ze zwiększeniem sprzedaży produktów wybranego producenta.
    \item Reguła biznesowa priorytetowo traktuje produkty posiadające wysokie stany magazynowe, zachęcając klienta do zakupu produktu, mimo, że istnieją produkty bardziej pasujące do bieżącego kontekstu.
\end{itemize}

\begin{figure}
	\centering
	\includegraphics[width=0.95\textwidth]{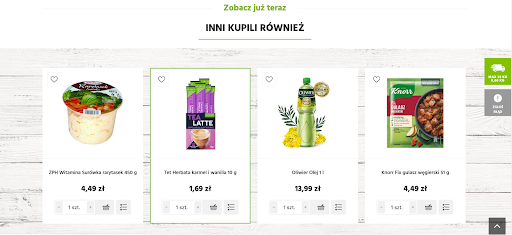}
	\caption{Przykładowa ramka rekomendacyjna.}
	\label{fig:fig1}
\end{figure}

Trzeba sobie wyraźnie powiedzieć, że silniki rekomendacyjne są bardzo wygodnym narzędziem manipulacji i stosunkowo łatwo jest wprowadzać do nich reguły, które mogą działać na niekorzyść klientów. Dodatkowym problemem jest to, że ramki rekomendacyjne mogą występować w sklepie internetowym w kontekstach, których odróżnianie może być dla klientów trudne. Przykładowo, ramka rekomendacyjna zatytułowana “\textit{Rekomendowane dla Ciebie}” może generować rekomendacje na podstawie wcześniejszej historii zakupów danego klienta, ale już ramka zatytułowana “\textit{Inni klienci kupili także}” (Rys.~\ref{fig:fig1}) do rekomendacji może wykorzystywać korelacje zakupowe wyznaczone na podstawie większej grupy klientów. Tu pojawia się pytanie: czy silnik rekomendacyjny wyszukał wcześniej klientów, którzy pod pewnym względem są podobni do bieżącego klienta i na tej próbce wyznaczył rekomendowane produkty? Jeśli tak, to należy zbadać, jakie cechy zostały wykorzystane do wyznaczenia tych “klientów najbardziej podobnych” --- może się okazać, że algorytm korzysta z cech chronionych (wiek, płeć), produkując rekomendacje obciążone specyficznymi stereotypami. A możemy mieć przecież jeszcze ramki rekomendacyjne na stronach kategorii produktowych, ramki rekomendacyjne w samym koszyku klienta (odpowiednik tzw. produktów impulsowych, czyli produktów oferowanych w bezpośrednim sąsiedztwie kas w tradycyjnych sklepach), czy ramki rekomendacyjne na stronie produktu. Każda z wyżej wymienionych ramek rekomendacyjnych może wykorzystywać inny algorytm wyznaczania rekomendacji i optymalizować różne funkcje celu. Skoro dyrektywa Omnibus wymaga wyjaśniania kryteriów sortowania wyników wyszukiwania, dlaczego nie wymaga wyjaśniania kryteriów stosowanych do budowania rekomendacji?

\section{Omnibus a silniki rekomendacji}

Wydaje się, że dyrektywa Omnibus pominęła istotny element wchodzący w skład większości współczesnych sklepów online: silniki rekomendacji. Rekomendacje w sklepie internetowym można postrzegać jako rodzaj skomplikowanego protokołu wyszukiwania, aczkolwiek jest on inicjowany przez sprzedawcę, a nie konsumenta. Ten niewidoczny lub ukryty proces wyszukiwania jest napędzany przez zaawansowane algorytmy zaprojektowane w celu przewidywania i dostosowywania się do zachowań, gustów i nawyków zakupowych klientów. Zasadniczo algorytm nieustannie przeszukuje dane, w tym historię zakupów klienta, informacje o profilu, historię przeglądarki i wszelkie aktualnie rozważane produkty. Elementy wyświetlane jako rekomendacje są zatem odpowiedziami systemu na te powtarzające się i samo-aktualizujące się parametry wyszukiwania. Celem jest rozszyfrowanie preferencji konsumenta i przedstawienie spersonalizowanego doświadczenia zakupowego, podobnego do wyników wyszukiwania na żądanie obejmujących produkty, którymi klient jest najprawdopodobniej zainteresowany.

Chociaż silniki rekomendacji mogą zapewnić spersonalizowane i usprawnione doświadczenie zakupowe, ich podstawowa moc może być niestety również wykorzystywana do nieuczciwych praktyk. Na przykład, systemy te mogą strategicznie umieszczać droższe produkty w rekomendacjach, odwracając uwagę użytkowników od opłacalnych wyborów, powodując tym samym, że konsumenci wydają więcej, niż pierwotnie zamierzali. Praktyka ta może powodować umieszczanie droższych produktów na wyższych pozycjach tylko dlatego, że generują one większe zyski, a nie ze względu na ich przydatność lub wartość dla klienta. Ponadto silniki te mogą tworzyć poczucie pilności, zachęcając użytkowników fałszywymi wskazówkami dotyczącymi popytu, takimi jak "Została tylko 1 sztuka" lub "20 osób patrzy na to w tej chwili", zmuszając w ten sposób użytkowników do podejmowania pochopnych decyzji zakupowych. Techniki takie jak te manipulują zachowaniem konsumentów, sprzyjają impulsywnym zakupom, a nie obiektywnemu, opartemu na potrzebach procesowi zakupu. Co gorsza, silniki rekomendacji mogą wykorzystywać dane użytkowników do tworzenia perswazyjnych narracji, na przykład sugerując wysokokaloryczne produkty osobom z zaburzeniami odżywiania lub agresywnie promując aplikacje hazardowe osobom podatnym. To tylko kilka sposobów, w jakie silniki rekomendacji mogą potencjalnie nadużywać swojego głębokiego wpływu na konsumentów online.

\subsection{Przejrzystość algorytmów i odpowiedzialność}

Aby wprowadzić wystarczającą ochronę klientów sklepów internetowych w odniesieniu do systemów rekomendacji, należy podjąć pewne kroki. Po pierwsze, kluczowe znaczenie ma demistyfikacja tak zwanej "czarnej skrzynki" silnika rekomendacji. Oznacza to otwarcie funkcjonowania algorytmów rekomendacji na kontrolę, co w konsekwencji zmniejsza niejasności wokół tych silników. Przejrzyste zaangażowanie i otwarte praktyki kodowania mogą przyczynić się do rzucenia światła na wewnętrzne działanie technologii. Po drugie, sklepy internetowe powinny zapewnić jasne i zrozumiałe wyjaśnienia dotyczące sposobu generowania rekomendacji. Można to osiągnąć za pomocą prostego języka i pomocy wizualnych, aby wyjaśnić, w jaki sposób dane użytkownika są wykorzystywane do generowania spersonalizowanych rekomendacji. Użytkownicy końcowi tych systemów powinni być w stanie łatwo zrozumieć czynniki wpływające na ich sugestie dotyczące produktów. Wreszcie, kluczowe znaczenie ma identyfikacja i łagodzenie potencjalnych uprzedzeń w algorytmach rekomendacji. Wymaga to dogłębnej analizy zaprojektowanych algorytmów w celu wykrycia wszelkich istniejących uprzedzeń, być może tych faworyzujących określone produkty, marki lub zachowania użytkowników. Po wykryciu takich uprzedzeń należy wyraźnie zająć się nimi i złagodzić je w celu opracowania bardziej sprawiedliwego i zrównoważonego systemu. To właśnie systematyczne podejście do przejrzystości, jasności i bezstronności zapewni klientom sklepów internetowych odpowiednią ochronę w kontaktach z systemami rekomendacji.

\subsection{Kontrola i wzmocnienie pozycji konsumenta}

Aby wzmocnić pozycję konsumentów, zmiany w silnikach rekomendacyjnych muszą promować przejrzystość i autonomię użytkowników. Dlatego umożliwienie użytkownikom zrozumienia i zarządzania preferencjami dotyczącymi rekomendacji jest koniecznością. Można to osiągnąć poprzez zapewnienie dostępnych, intuicyjnych narzędzi i ustawień, które pozwalają użytkownikom na tworzenie własnych rekomendacji, dając im tym samym kontrolę nad ich zakupami online. Ponadto należy wdrożyć solidne mechanizmy rezygnacji i dostępu do danych. Klienci powinni mieć prawo do łatwej rezygnacji lub ograniczenia danych wykorzystywanych przez systemy rekomendacji, a to wymaga jasnych i łatwo dostępnych narzędzi do przeglądania, modyfikowania lub usuwania. Wreszcie, wspieranie zaufania do spersonalizowanych rekomendacji powinno być głównym celem. Wiąże się to z zobowiązaniem do przejrzystości - jasnego komunikowania, w jaki sposób i dlaczego niektóre produkty są zalecane, oraz wykazywania ciągłego zaangażowania w uczciwość i dokładność tych sugestii. Łącznie zmiany te oddają kontrolę z powrotem w ręce konsumentów, pielęgnują zaufanie i ostatecznie poprawiają ogólne wrażenia użytkowników z silników rekomendacji w sklepach internetowych.

\section{Wnioski}

Kontrolowanie silników rekomendacji online w sklepach internetowych wiąże się z szeregiem wyzwań, które mogą być dość zniechęcające. Ewoluujący krajobraz systemów rekomendacji dodatkowo komplikuje sprawę. Innowacje w tym obszarze są ciągłe, a nowe technologie pojawiają się w szybkim tempie, co może sprawić, że istniejące systemy staną się przestarzałe lub nieaktualne. Dostosowanie się do tych nowych technologii i trendów wymaga ciągłej czujności i chęci adaptacji. Wreszcie, kluczowym wyzwaniem jest znalezienie równowagi. Chociaż innowacje są kluczem do utrzymania zainteresowania konsumentów i zachowania konkurencyjności, nie mogą one odbywać się kosztem ochrony konsumentów. Istnieje skomplikowane wyzwanie polegające na opracowaniu algorytmów rekomendacji, które nie tylko przedstawiają odpowiednie sugestie, ale także szanują prywatność danych konsumentów i zachowują etyczne stanowisko. Każde z tych wyzwań stanowi własną przeszkodę, wymagającą pomysłowości i elastyczności, aby skutecznie kontrolować silniki rekomendacji online w dzisiejszym dynamicznym środowisku e-commerce.

Niedawne wejście w życie dyrektywy Omnibus obiecuje znacząco zmienić krajobraz zakupów online. Działając jako katalizator zmian, przepisy te mają na celu modernizację prawa konsumenckiego i skuteczne wypełnienie luki między szybką ewolucją rynków cyfrowych a istniejącą ochroną prawną konsumentów. Oczekuje się, że dyrektywa zapewni większą przejrzystość platform internetowych i zajmie się nieuczciwymi praktykami związanymi ze sprzedażą online. Jednym z kluczowych przepisów jest obowiązek jasnego informowania z góry o tym, w jaki sposób niejawne czynniki wpływają na ranking produktów i czy dane osobowe są wykorzystywane do celów komercyjnych. Co więcej, internetowe platformy handlowe będą musiały informować konsumentów o tożsamości głównego sprzedawcy, aby odróżnić sprzedawców prywatnych od profesjonalnych. Wdrażając takie solidne zmiany, dyrektywa Omnibus zapoczątkuje istotną zmianę w cyfrowej ochronie konsumentów, wspierając bezpieczniejsze i bardziej odpowiedzialne zakupy online dla obywateli UE.

Naszym zdaniem konieczne jest rozszerzenie dyrektywy Omnibus w taki sposób, aby stosowne regulacje objęły także treści generowane przez systemy rekomendacyjne. Systemy te, powszechnie stosowane przez szeroką gamę platform internetowych, znacząco wpływają na wybory i zachowania użytkowników. Chociaż ich korzystny wpływ na usprawnienie doświadczenia użytkownika jest niezaprzeczalny, nie możemy przymykać oczu na potencjalne ryzyko związane z brakiem przejrzystości i odpowiedzialności. Aby zapobiec wszelkim szkodliwym skutkom, należy wdrożyć dwa kluczowe elementy. Po pierwsze, każdy system rekomendacji powinien zostać poddany ocenie etycznej, aby upewnić się, że jego funkcjonalność nie prowadzi do potencjalnie szkodliwych rekomendacji. Taka kontrola mogłaby pomóc złagodzić uprzedzenia i chronić wrażliwe segmenty populacji. Audyt etyczny systemu rekomendacyjnego może zostać przeprowadzony w postaci ustandaryzowanego benchmarku albo udokumentowanego badania wewnętrznego na domenowym zbiorze danych. Po drugie, kryteria leżące u podstaw algorytmów rekomendacji powinny być wyraźnie komunikowane użytkownikom, odzwierciedlając wymogi przejrzystości dotyczące rankingu wyników wyszukiwania. W ten sposób użytkownicy mogą lepiej zrozumieć, dlaczego wydawane są określone rekomendacje, co sprzyja bardziej zaufanemu i uczciwemu środowisku internetowemu. Ma to szczególne znaczenie w sytuacji, gdy niezwykle podobny interfejs (ramka rekomendacyjna) zawiera propozycje wygenerowane przez zupełnie różne kryteria.

Chociaż dyrektywa Omnibus jest niezaprzeczalnie ogromnym krokiem naprzód w zakresie ochrony konsumentów internetowych, jednym z jej krytycznych niedociągnięć jest brak wyraźnych przepisów regulujących potencjalne nieuczciwe praktyki związane z silnikami rekomendacji. Te inteligentne systemy mają znaczący wpływ na zachowania konsumentów i tendencje zakupowe, ale ich działania pozostają w dużej mierze niekontrolowane przez standardy określone w dyrektywie. Dzięki swojej zdolności do subtelnej zmiany dynamiki handlu internetowego poprzez manipulacyjne sugestie i potencjalnie nieuczciwe rekomendacje, takie silniki mogą utrwalać kwestie, które dyrektywa ma na celu unicestwić. Dlatego też kluczowe znaczenie dla przyszłych wersji dyrektywy ma rozszerzenie jej zakresu na te systemy rekomendacji. Ustanowienie przepisów dotyczących ich konstrukcji i funkcjonalności nie tylko zapewniłoby bezpieczną przestrzeń zakupową dla konsumenta, ale także ugruntowałoby kompleksowe ramy ochrony na szybko ewoluującym rynku internetowym.

\appendix

\textbf{Finansowanie}: Inteligentny System Rekomendacji został opracowany przez specjalistów Ad First sp. z o.o. przy współpracy z naukowcami z Wydziału Informatyki i Telekomunikacji Politechniki Poznańskiej w ramach projektu  pn.: „\emph{Opracowanie skalowalnego prototypu inteligentnego systemu personalizacji doświadczeń użytkowników dla branży e-Commerce z wykorzystaniem uczenia maszynowego oraz semantycznej analizy danych heterogenicznych}”. Projekt został zrealizowany ze środków Unii Europejskiej w ramach Działania 1.2 Wielkopolskiego Regionalnego Programu Operacyjnego.

\end{document}